\documentclass[12pt, draftclsnofoot,onecolumn]{IEEEtran}

\usepackage{graphicx,subcaption,xcolor,setspace,amsthm,cite,booktabs,epstopdf,dblfloatfix,blindtext,amsthm}
\usepackage[hyphens]{url}
\usepackage{algorithm,algcompatible,upgreek,acro}
\usepackage{mathtools, cuted}
\usepackage{balance,makecell}

\usepackage[binary-units,detect-weight=true, detect-family=true]{siunitx}
\DeclareSIUnit \bitspersecond {bps}
\usepackage{amsmath,amssymb}

\newcommand{\bq}{{\mathbf{q}}}

\newcommand{\bzero}{{\mathbf{0}}}

\newcommand{\diag}{{\rm diag}}

\makeatletter
\def\munderbar#1{\underline{\sbox\tw@{$#1$}\dp\tw@\z@\box\tw@}}
\makeatother
\begin{document}
\newtheorem{theorem}{Theorem}
\newtheorem{corollary}{Corollary}
\newtheorem{guess}{Proposition}
\newtheorem{guess1}{Lemma}
\newcommand{\lam}{\mbox{\small{\boldmath $\lambda$}}}
\newcommand{\Lam}{\mbox{\boldmath $\Lambda$}}
\newcommand{\sig}{\mbox{\boldmath $\sigma$}}
\newcommand{\Sig}{\mbox{\boldmath $\Sigma$}}
\def\diag{{\mathbf diag}}
\newcommand{\BMlambda}{\mbox{\small{\boldmath $\lambda$}}}
\newcommand{\BMLambda}{\mbox{\boldmath $\Lambda$}}
\newcommand{\BMd}{\mbox{\small{\boldmath $d$}}}
\newcommand{\BMu}{\mbox{\small{\boldmath $u$}}}
\newcommand{\BMw}{\mbox{\small{\boldmath $w$}}}
\newcommand{\BMn}{\mbox{\small{\boldmath $n$}}}
\newcommand{\BMx}{\mbox{\small{\boldmath $x$}}}
\newcommand{\BMy}{\mbox{\small{\boldmath $y$}}}
\newcommand{\BMR}{\mbox{\small{\boldmath $R$}}}
\newcommand{\EE}{\mathsf{E}}
\newcommand{\herm}{{\sf H}}

\bstctlcite{IEEEmax3beforeetal}
\title{Mean-square Analysis of the NLMS Algorithm}
\author{Tareq Y. Al-Naffouri, Muhammad Moinuddin, and Anum Ali 
\thanks{T. Y. Al-Naffouri is with the Computer, Electrical and Mathematical Sciences and Engineering Division, King Abdullah University of Science and Technology, Thuwal 23955-6900, Saudi Arabia \mbox{(e-mail: tareq.alnaffouri@kaust.edu.sa)}.}
\thanks{M. Moinuddin is with the Department of Electrical and Computer
Engineering, King Abdulaziz University, Jeddah 21589, Saudi Arabia, and also
with the Center of Excellence in Intelligent Engineering Systems (CEIES),
King Abdulaziz University, Jeddah 21589, Saudi Arabia \mbox{(e-mail: mmsansari@
kau.edu.sa)}.}
\thanks{A. Ali is with the Standards and Mobility Innovation Laboratory, Samsung Research America, Plano, TX 75023, USA \mbox{(e-mail: anum.ali@samsung.com)}.}
}

\maketitle

\begin{abstract}
This work presents a novel approach to the mean-square analysis of the normalized least mean squares (NLMS) algorithm for circular complex colored Gaussian inputs. The analysis is based on the derivation of a closed-form expression for the Cumulative Distribution Function (CDF) of random variables
of the form $(||{\bf u}||_{{\bf D}_1}^2)(||{\bf u}||_{{\bf D}_2}^2)^{-1}$ where ${\bf u}$ is an isotropic vector and ${\bf D}_1$ and ${\bf D}_2$ are diagonal matrices and using that to derive some moments of these variables. These moments in turn completely characterize
the mean-square behavior of the NLMS algorithm in explicit closed-form expressions. Specifically, the transient, steady-state, and tracking mean-square behavior of the NLMS algorithm is studied. 
\end{abstract}


\section{Introduction}
\label{sec:intro}
The most widely used algorithm for adaptive filters is the Least Mean Squares (LMS) \cite{sayed} algorithm which
provides a solution to the optimal Weiner Filter criterion minimizing the mean square value of the error in a stochastic
approximation sense. Since the LMS algorithm belongs to such a class of gradient-type algorithms whose convergence properties
depend on the input correlation, it inherits slow convergence, especially when operating on highly correlated signals like speech. This dependency is removed by the input normalization in the NLMS algorithm \cite{NLMS} which results in a great improvement in convergence compared to the LMS algorithm \cite{sayed}. However, the normalization complicates the performance analysis of the
algorithm.

Many researchers have attempted to study the performance of the NLMS~\cite{TranDataNon,sayed,Douglas,Barrault,Lobato,CostaBermudaz,Ali,Slock} and to understand why it has superior performance over that of the LMS. In \cite{TranDataNon}, closed-form expressions for the transient analysis and the steady-state
mean-square error (MSE) of the NLMS algorithm are developed but these expressions are in terms of multidimensional moments which
\cite{TranDataNon} falls short of evaluation. Other prior work has attempted to evaluate these moments but the corresponding
analyses do not result in closed form performance expressions \cite{sayed,TranDataNon,Douglas}, or rely on strong assumptions.
Examples of these assumptions include the separation principle \cite{Barrault,Lobato}, approximations \cite{CostaBermudaz}, white input \cite{Douglas,CostaBermudaz,Slock}, specific structure of input regressor’s distribution \cite{Barrault,Tarrab}, small step size \cite{Barrault}, long filters \cite{CostaBermudaz}, approximate solutions using Abelian integrals \cite{Barrault} and partial evaluation of moments \cite{NLMSmean}. 

In this work, we obtain the closed form expressions for the required multidimensional input moments. Our approach is based on evaluating the CDF and the moments of random variables of the form\footnote{For any matrix ${\bf D}$, the quadratic form $||{\bf a}||_{{\bf D}}^2$ is defined as  $||{\bf a}||_{{\bf D}}^2\stackrel{\triangle}{=}{\bf a}^{*}{\bf D}{\bf a}$.} $\xi\left({\bf u},{\bf D}_1,{\bf D}_2\right)=
(||{\bf u}||_{{\bf D}_1}^2)(||{\bf u}||_{{\bf D}_2}^2)^{-1}$  where ${\bf u}$ is a white Gaussian vector and ${\bf D}_1$ and ${\bf D}_2$ are diagonal matrices, by expressing them as ratios of quadratic forms $(||{\mbox{\boldmath $\phi$}}||_{{\bf D}_1}^2)(||{\mbox{\boldmath $\phi$}}||_{{\bf D}_2}^2)^{-1}$ in isotropic random vector ${\mbox{\boldmath $\phi$}}$. We then use complex integration to evaluate the CDF and the moments of these ratios. 

The closed-form expressions obtained in this work permit the transient, steady-state, and tracking mean-square analyses of the NLMS algorithm in a (non) stationary environment for colored circular Gaussian inputs. Unlike existing work, our approach neither uses the separation principle nor the small step-size assumption. As a by-product of our approach, we demonstrate how to statistically characterize indefinite quadratic forms in isotropic random vectors using complex integration. These quantities are of vital importance in signal processing, communications, and information theory and researchers have resorted to non-traditional techniques to study these forms \cite{Mathai,Johnson,isotropic}. To the best of our knowledge, this is the first work that provides the closed-form expressions for the performance of the NLMS. 

The paper is organized as follows. Following this introduction, we carry out the transient, tracking, and steady-state mean-square analyses of the NLMS algorithm in Section~\ref{performance} and point out the moments that need to be evaluated to express the performance in closed form. In Section~\ref{approach}, the approach of our work is outlined. Derivation for the CDF of the random variable of the form $(||{\mbox{\boldmath $\phi$}}||_{{\bf D}_1}^2)(||{\mbox{\boldmath $\phi$}}||_{{\bf D}_2}^2)^{-1}$ as well as the derivation of its moments are carried out in Section IV. In Section~\ref{sim}, we present the simulation results to verify the analysis, followed by our conclusion in Section~\ref{con}.
\section{Performance Analysis of the NLMS Algorithm}\label{performance}

\subsection{System Model}\label{sysmodel}
Given a sequence of desired response $\{d_i\}$ and a sequence of regressor (row) vectors $\{{\bf u}_i\}$, an adaptive filter generates a weight vector ${\bf w}_{i}$
at each instant so that ${\bf u}_i {\bf w}_i$ is a good estimate of $d_i$. In the NLMS algorithm, the weight vector starting from ${\bf w}_{-1}=0$ is updated according to 
\begin{equation}\label{eqNLMS}
{\bf w}_{i}={\bf w}_{i-1}+\mu \frac{{\bf u}_i^{*}}{||{\bf u}_i||^2}e_i,\hspace{1cm}i\geq 0,
\end{equation}
where $e_i=d_i-{\bf u}_i {\bf w}_{i-1}$ is the estimation error and $\mu$ is the step-size. In performing the analysis of the NLMS algorithm, we consider the time variant case\footnote{Note that time invariant case is a special scenario of time variant case which is obtained by setting ${\bf q}_i=0$ for all $i$.} and hence, $d_i$ is generated by the time-variant system identification model $d_i={\bf u}_i {\bf w}_i^o+v_i$, where $v_i$ is a zero mean i.i.d. noise sequence with variance $\sigma_v^2$ and ${\bf w}_i^o$ varies according to
\begin{equation}\label{TV2}
{\bf w}_i^o={\bf w}_{i-1}^o+{\bf q}_i,
\end{equation}
and where ${\bf q}_i$ is assumed to be i.i.d. colored circular complex Gaussian ${\bf q}_i \sim {\mathcal{C}}{\mathcal{N}}~({\bf 0},{\bf R}_q)$. The ${\bf w}_i$ generated by the adaptive filter attempts to track the time variant system ${\bf w}_i^o$. Let ${\tilde {\bf w}}_{i}={\bf w}_i^o-{\bf w}_i$ denote the weight error vector, then
\begin{equation}\label{eq2new}
e_i={\bf u}_i ({\bf w}_i^o-{\bf w}_{i-1})+v_i={\bf u}_i {\tilde {\bf w}}_{i-1}+v_i+{\bf u}_i{\bf q}_i, 
\end{equation}
and, by subtracting ${\bf w}_i$ from both sides of (\ref{TV2}) and substituting the value of ${\bf w}_i$ from (\ref{eqNLMS}),
the NLMS update can be equivalently written as
\begin{equation}\label{eq3}
\tilde{{\bf w}}_{i}=\tilde{{\bf w}}_{i-1}-\mu \frac{{\bf u}_i^{*}}{||{\bf u}_i||^2}e_i+{\bf q}_i,\hspace{1cm}i\geq 0.
\end{equation}
We assume that the sequences $\{v_i\},~\{{\bf u}_i\}$, and $\{{\bf q}_i\}$ are mutually independent which is well justified in real practice. In addition, we will restrict our attention in this work to circularly symmetric Gaussian inputs, i.e. ${\bf u}_i \sim {\mathcal{C}}{\mathcal{N}}~({\bf 0},{\bf R}_u)$. We also assume that the input vectors are independent but their entries are correlated. For the sake of mean-square analysis, the autocorrelation matrix ${\bf R}_u$ can be assumed to be diagonal, that is, ${\bf R}_u=\Lam= \mbox{diag}(\lambda_1,\lambda_2,\cdots,\lambda_M)$, without loss of generality \cite{TranDataNon,Douglas}.

\subsection{Mean Square Performance of the NLMS Algorithm}\label{MSEstationary}
Starting from the recursion (\ref{eq3}) and using an approach similar to that of \cite{sayed,TranDataNon}, we can show that the weighted variance relation of the NLMS weight error vector results in\footnote{Here, ${\bf u}_i$ has diagonal correlation matrix $\Lam$ and ${\bf q}_i$ has rotated correlation matrix ${\bf R}_{\bar{q}}$.}
\begin{equation}\label{wtdvar11}
E[||{\tilde {\bf w}}_{i}||^2_{\sig}]=E[||{\tilde {\bf w}}_{i-1}||^2_{{\bf
F}\sig}]+\mu^2\sigma_v^2E\left[\frac{||{\bf u}_i||^2_{\sig}}{(||{\bf u}_i||^2)^2}\right]+E[||{\bf q}_i||^2_{{\bf F}\sig}],
\end{equation}
where $\sig$ is an $M\times 1$ parameter weight vector that can provide different performance measures by choosing its value appropriately. The matrix ${\bf F}$ is defined as
\begin{equation}\label{Fdef}
{\bf F}\mathop{=}^{\triangle}{\bf I}-\mu{\bf A}+\mu^2{\bf B},
\end{equation}
where ${\bf I}$ is the identity matrix and ${\bf A}$ and ${\bf B}$ are the multidimensional input moment matrices
\begin{equation}
{\bf A}\mathop{=}^{\triangle}2E\left[\frac{{\bf u}^{*}_i{\bf u}_i}{||{\bf u}_i||^2}\right],
~~{\bf B}\mathop{=}^{\triangle}E\left[\frac{({\bf u}_i^{*} {\bf u}_i)^T\odot({\bf u}_i^{*} {\bf u}_i)}{(||{\bf u}_i||^2)^2}\right],\label{ABdef}
\end{equation}
where the notation $\odot$ denotes an element-by-element (Hadamard) product. Equation (\ref{wtdvar11}) can be used to describe
both the transient and steady-state performance of the NLMS algorithm. In order to derive an expression for the steady-state EMSE ($\zeta$), we evaluate (\ref{wtdvar11}) by taking the limit of both sides of (\ref{wtdvar11}) as $i\rightarrow\infty$ and utilizing the fact that the terms $E[||{\tilde {\bf w}}_{i}||^2_{\sig}]$ and $E[||{\tilde {\bf w}}_{i-1}||^2_{{\bf F}\sig}]$ can be combined to $E[||{\tilde {\bf w}}_{\infty}||^2_{(\bf{I}-{\bf F})\sig}]$ at $i\rightarrow\infty$. Thus, by setting $\sig=(\bf{I}-{\bf F})^{-1}\lam$ (where $\lam$ is a vector defined as $\lam=\mbox{diag}(\Lam)$), we can show that the steady-state EMSE of the NLMS is given by
\begin{equation}\label{steadyEMSE}
\zeta_{track}=\mu^2\sigma_v^2\mbox{Tr}({\bf C}{\bf G})+\mbox{Tr}({\bf R}_{\bar{q}}{\bf G}_{\bar{q}}),
\end{equation}
where ${\bf C}$ is the multidimensional moment 
\begin{equation}
{\bf C}\mathop{=}^{\triangle}E\left[\frac{{\bf u}^{*}_i{\bf u}_i}{(||{\bf u}_i||^2)^2}\right],\label{Cdef}
\end{equation}
and ${\bf G}$ and ${\bf G}_{\bar{q}}$ are diagonal matrices such that $\mbox{diag}({\bf G})=(\bf{I}-{\bf F})^{-1}\lam$ and $\mbox{diag}({\bf G}_q)={\bf F}(\bf{I}-{\bf F})^{-1}\lam$. It can be noted that the steady-state EMSE of the time invariant case can be directly deduced from (\ref{steadyEMSE}) by setting ${\bf q}_i$ to zero.

A summary of the various mean-square performance measures of the NLMS algorithm obtained from (\ref{wtdvar11}) is shown in Table \ref{tab1}.

\begin{table}[h]
\caption{Summary of analytical results for various performance measures of the NLMS algorithm.} \label{tab1}
\begin{center} {
\begin{tabular}{|l|l|l|}
\hline Performance Measure & Analytical Result & Moments Required
\\ \hline EMSE Learning Curve & $E[||{\tilde {\bf w}}_i||^2_{\lam}]=E[||{\bf w}^o||^2_{{\bf F}^i\lam}]+\mu^2\sigma_v^2E\left[\frac{||{\bf
u}_i||^2_{({\bf I}+{\bf F}+\cdots+{\bf F}^{i-1})\lam}}{(||{\bf u}_i||^2)^2}\right]$ & \hspace{.8cm}${\bf A},{\bf B},{\bf C}$
\\ \hline MSD Learning Curve & $E[||{\tilde {\bf w}}_i||^2_{\bf 1}]=E[||{\bf w}^o||^2_{{\bf F}^i{\bf 1}}]+\mu^2\sigma_v^2E\left[\frac{||{\bf
u}_i||^2_{({\bf I}+{\bf F}+\cdots+{\bf F}^{i-1}){\bf 1}}}{(||{\bf u}_i||^2)^2}\right]$ & \hspace{.8cm}${\bf A},{\bf B},{\bf C}$
\\ \hline Mean Stability & $0<\mu<2/\lambda_{max}(\frac{1}{2}{\bf A})$ & \hspace{1.3cm}${\bf A}$
\\ \hline Mean Square Stability & $0<\mu<1/\lambda_{max}({\bf A}^{-1}{\bf B})$ & \hspace{1.1cm}${\bf A}, {\bf B}$
\\ \hline Steady-state EMSE & $\mu^2\sigma_v^2E\left[\frac{||{\bf u}_i||^2_{(\bf{I}-{\bf F})^{-1}\lam}}{(||{\bf
u}_i||^2)^2}\right]$ & \hspace{.8cm}${\bf A},{\bf B},{\bf C}$
\\ \hline Steady-state MSD & $\mu^2\sigma_v^2E\left[\frac{||{\bf u}_i||^2_{(\bf{I}-{\bf F})^{-1}{\bf 1}}}{(||{\bf
u}_i||^2)^2}\right]$ & \hspace{.8cm}${\bf A},{\bf B},{\bf C}$
\\ \hline Tracking EMSE & $\mu^2\sigma_v^2E\left[\frac{||{\bf u}_i||^2_{(\bf{I}-{\bf F})^{-1}\lam}}{(||{\bf
u}_i||^2)^2}\right]+E\left[||{\bf q}_i||^2_{{\bf F}(\bf{I}-{\bf F})^{-1}\lam}\right]$ & \hspace{.8cm}${\bf A},{\bf B},{\bf C}$
\\ \hline
\end{tabular}}
\end{center}
\end{table}
\section{Our Methodology}\label{approach}
By inspecting Table \ref{tab1}, we see that all the MSE performance measures are completely determined by the moment matrix ${\bf F}$ (and hence the moment matrices ${\bf A}$
and ${\bf B}$) and by the moment matrix\footnote{Table \ref{tab1} contains entries of the form $E\left[\frac{||{\bf u}_i||^2_{{\bf G}}}{(||{\bf u}_i||^2)^2}\right]$ for some weighting matrix ${\bf G}$. Knowing that ${\bf C}$ is a diagonal matrix (see Table \ref{tab3}), it is easy to show that $E\left[\frac{||{\bf u}_i||^2_{{\bf G}}}{(||{\bf u}_i||^2)^2}\right]=\mbox{Tr}({\bf C}{\bf G})$} ${\bf C}$. In order to evaluate these moment matrices, we rewrite them as follows
\begin{equation}
{\bf A}\mathop=\Lam \bar{\bf A},~~{\bf B}=\Lam \bar{\bf B} \Lam,~~\mbox{and}~~{\bf C}=\Lam \bar{\bf C},\label{ABCdef2}
\end{equation}
where $\bar{\bf A}$, $\bar{\bf B}$, and $\bar{\bf C}$ stand for the ``whitened" versions of the matrices ${\bf A}$, ${\bf B}$, and ${\bf C}$, respectively, i.e.,
\begin{equation}
\bar{\bf A}\mathop{=}^{\triangle}2E\left[\frac{\bar{{\bf u}}^{*}_i\bar{{\bf u}}_i}{||{\bf u}_i||^2}\right],~~\bar{\bf B}\mathop{=}^{\triangle} E\left[\frac{(\bar{{\bf u}}_i^{*} \bar{{\bf u}}_i)^T\odot(\bar{{\bf u}}_i^{*} \bar{{\bf u}}_i)}{(||{\bf u}_i||^2)^2}\right],
~~\mbox{and}~~\bar{\bf C}\mathop{=}^{\triangle}E\left[\frac{\bar{{\bf u}}^{*}_i\bar{{\bf u}}_i}{(||{\bf u}_i||^2)^2}\right].\label{D123def}
\end{equation}
Here $\bar{u}_i(k)$ is the whitened version of $u_i(k)$, i.e. $u_i(k)=\sqrt{\lambda_k}\bar{u}_i(k)$, where $\lambda_k$ are the eigenvalues of the autocorrelation of the input sequence making the vector $\bar{{\bf u}}_i$ of identity correlation matrix, i.e., $\bar{\bf u}_i \sim {\mathcal{C}}{\mathcal{N}}~({\bf 0},{\bf I})$. Note that $||{\bf u}_i||^2=||\bar{\bf u}_i||_{\Lambda}^2$. Thus,
from (\ref{D123def}), it can be easily deduced that the entries of the moment matrices $\bar{\bf A}$, $\bar{\bf B}$, and $\bar{\bf C}$ are completely determined by the expectation of the following
\begin{equation}\label{randv}
\frac{|\bar{u}(k)|^2}{||\bar{\bf u}||_{\Lambda}^2},~~\frac{|\bar{u}(k)|^2|\bar{u}(\bar{k})|^2}{(||\bar{\bf u}||_{\Lambda}^2)^2},~~\frac{|\bar{u}(k)|^2}{(||\bar{\bf u}||_{\Lambda}^2)^2},~~~(\mbox{with}~ k\neq \bar{k}),
\end{equation}
where $k$ and $\bar{k}$ denote the $k^{th}$ and $\bar{k}^{th}$ distinct positions of the regressor elements in the input vector
$\bar{{\bf u}}$. Here we have dropped the subscript $i$ for convenience of notation. To evaluate these moments, we will express
them in terms of moments of ratios of {\it quadratic forms} as it is relatively easy to characterize these ratios statistically (i.e.
to determine their pdf/CDF). The first random variable
\begin{equation}\label{sk}
s_k\stackrel{\triangle}{=}\frac{|\bar{u}(k)|^2}{||\bar{\bf u}||_{\Lambda}^2},
\end{equation}
is already a ratio of quadratic forms. The second random variable in (\ref{randv}) can be expressed with the aid of the binomial
expansion as
\begin{equation}
\frac{|\bar{u}(k)|^2|\bar{u}(\bar{k})|^2}{(||\bar{\bf u}||_{\Lambda}^2)^2}=\frac{1}{2}s_{k\bar{k}}^2-\frac{1}{2}\frac{\lambda_k}{\lambda_{\bar{k}}}s_{k}^2-\frac{1}{2}\frac{\lambda_{\bar{k}}}{\lambda_k}s_{\bar{k}}^2,
\end{equation}
where $s_{k\bar{k}}$ is defined as
\begin{equation}\label{skkbar}
s_{k\bar{k}}\stackrel{\triangle}{=}\frac{\sqrt{\frac{\lambda_k}{\lambda_{\bar{k}}}}|\bar{u}(k)|^2+\sqrt{\frac{\lambda_{\bar{k}}}{\lambda_k}}|\bar{u}(\bar{k})|^2}{||\bar{\bf u}||_{\Lambda}^2},
\end{equation}
and so~\eqref{skkbar} is a sum of two ratios of quadratic forms. Finally, in order to evaluate the moment of third random variable in (\ref{randv}), we define the following two new variables
\begin{equation}\label{zrdef}
z_k\stackrel{\triangle}{=}\frac{|\bar{u}(k)|^2+1}{||\bar{\bf u}||_{\Lambda}^2},~~~\mbox{and}~~r\stackrel{\triangle}{=}\frac{1}{||\bar{\bf u}||_{\Lambda}^2}.
\end{equation}
Thus, the required random variable $\frac{|\bar{u}_k|^2}{(||\bar{\bf u}||_{\Lambda}^2)^2}$ can be obtained using binomial expansion as follows:
\begin{equation}
\frac{|\bar{u}(k)|^2}{(||\bar{\bf
u}||_{\Lambda}^2)^2}=\frac{1}{2}\left(\frac{|\bar{u}(k)|^2+1}{||\bar{\bf u}||_{\Lambda}^2}\right)^2-\frac{1}{2}\left(\frac{|\bar{u}(k)|^2}{||\bar{\bf u}||_{\Lambda}^2}\right)^2-\frac{1}{2}\left(\frac{1}{||\bar{\bf u}||_{\Lambda}^2}\right)^2=\frac{1}{2}z_k^2-\frac{1}{2}s_k^2-\frac{1}{2}r^2. \label{eqzk}
\end{equation}
Thus, all required moments are expressed as function of the moments of $s_k$, $s_{k\bar{k}}$, $r$, and $z_k$ as in Table
\ref{tab2}. With this, the entries of the matrices $\bar{\bf A}$, $\bar{\bf B}$, and $\bar{\bf C}$ can be expressed
directly in terms of the moments of these random variables as shown in Table \ref{tab3}. In the remaining part of the paper, we
derive the CDF's of $s_k$, $s_{k\bar{k}}$, $r$, and $z_k$ and subsequently use them to derive the moments listed in Table
\ref{tab3}.

\begin{table}[h]
\caption{Required moments and their relations to $s_k$, $s_{k\bar{k}}$, $r$, and $z_k$.}
\begin{center}
\begin{tabular}{|l|l|}
\hline Required Moments & \hspace{.5cm} Required Expressions
\\ \hline \hspace{.5cm}$E\left[\frac{|\bar{u}(k)|^2}{||\bar{\bf u}||_{\Lambda}^2}\right]$ & \hspace{1.4cm}$E\left[s_k\right]$
\\ \hline \hspace{.3cm}$E\left[\frac{|\bar{u}(k)|^2|\bar{u}(\bar{k})|^2}{(||\bar{\bf u}||_{\Lambda}^2)^2}\right]$ &
$\frac{1}{2}E\left[s_{k\bar{k}}^2\right]-\frac{1}{2}\frac{\lambda_k}{\lambda_{\bar{k}}}E\left[s_k^2\right]-\frac{1}{2}\frac{\lambda_{\bar{k}}}{\lambda_k}E\left[s_{\bar{k}}^2\right]$
\\ \hline \hspace{.3cm}$E\left[\left(\frac{1}{||\bar{\bf u}||_{\Lambda}^2}\right)^2\right]$ & \hspace{1.4cm}$E\left[r^2\right]$
\\ \hline \hspace{.2cm}$E\left[\left(\frac{|\bar{u}(k)|^2+1}{||\bar{\bf
u}||_{\Lambda}^2}\right)^2\right]$ & \hspace{1.4cm}$E\left[z_k^2\right]$
\\ \hline \hspace{.4cm}$E\left[\frac{|\bar{u}(k)|^2}{(||\bar{\bf u}||_{\Lambda}^2)^2}\right]$ &
\hspace{.2cm}$\frac{1}{2}E\left[z_k^2\right]-\frac{1}{2}E\left[s_k^2\right]-\frac{1}{2}E\left[r^2\right]$
\\ \hline
\end{tabular}\label{tab2}
\end{center}
\end{table}
\begin{table}
\begin{center}
\caption{Relation of entries of moment matrices $\bar{\bf A}$, $\bar{\bf B}$, and $\bar{\bf C}$ to $s_k$, $s_{k\bar{k}}$, $r$, and $z_k$.}
\begin{tabular}{|l|l|}
\hline Matrix Entries & \hspace{1.5cm}Required Moments
\\ \hline \hspace{.5cm}$\bar{\bf A}(k,k)$ & \hspace{1.8cm}$2E\left[s_k\right]$
\\ \hline \hspace{.5cm}$\bar{\bf A}(k,\overline{k})$ & \hspace{2.2cm}$0$
\\ \hline \hspace{.5cm}$\bar{\bf B}(k,k)$ & \hspace{1.8cm}$E\left[s_{k}^2\right]$
\\ \hline \hspace{.5cm}$\bar{\bf B}(k,\overline{k})$ &  $\frac{1}{2}E\left[s_{k\bar{k}}^2\right]-\frac{1}{2}\frac{\lambda_k}{\lambda_{\bar{k}}}E\left[s_k^2\right]-\frac{1}{2}\frac{\lambda_{\bar{k}}}{\lambda_k}E\left[s_{\bar{k}}^2\right]$
\\ \hline \hspace{.5cm}$\bar{\bf C}(k,k)$ & $\frac{1}{2}E\left[z_k^2\right]-\frac{1}{2}E\left[s_k^2\right]-\frac{1}{2}E\left[r^2\right]$
\\ \hline \hspace{.5cm}$\bar{\bf C}(k,\overline{k})$ & \hspace{2.2cm}$0$
\\ \hline
\end{tabular}\label{tab3}
\end{center}
\end{table}
\section{Evaluation of the CDF and Moments}
In this section, we derive the CDF and moments for all the required variables defined in previous section.
\subsection{The CDF of $s_k$ and $s_{k\bar{k}}$}\label{cdfskkbar}
The key to evaluating the CDF of $s_{k\bar{k}}$ is to first define
$s_{k\bar{k}}$ in terms of isotropic random variables. By using the definition
${\mbox{\boldmath $\phi$}}=\left[\frac{\bar{u}(1)}{\|\bar{u}\|},~\frac{\bar{u}(2)}{\|\bar{u}\|},\cdots,\frac{\bar{u}(M)}{\|\bar{u}\|}\right]$,
we can rewrite $s_{k\bar{k}}$ as
\begin{equation}\label{skkbarrep}
s_{k\bar{k}}=\frac{\sqrt{\frac{\lambda_k}{\lambda_{\bar{k}}}}|\bar{u}(k)|^2+\sqrt{\frac{\lambda_{\bar{k}}}{\lambda_k}}|\bar{u}(\bar{k})|^2}
{\sum_{i=1}^{M}\lambda_i|\bar{u}(i)|^2}=\frac{\|{\mbox{\boldmath $\phi$}}\|^2_{{\bf \Sigma}_{k\bar{k}}}}{\|{\mbox{\boldmath $\phi$}}\|^2_{\bf \Lambda}},~~~k\neq \bar{k},
\end{equation}
where ${\bf \Sigma}_{k\bar{k}}$ is an $M\times M$ matrix with all elements set to zero except the $k^{th}$ and $\bar{k}^{th}$ elements
on the main diagonal which are set to $\sqrt{\frac{\lambda_k}{\lambda_{\bar{k}}}}$ and $\sqrt{\frac{\lambda_{\bar{k}}}{\lambda_k}}$,
respectively. The random vector ${\mbox{\boldmath $\phi$}}$ in (\ref{skkbarrep}) is known as an isotropic random vector \cite{isotropic} and has the pdf
\begin{equation}
p(\phi)=\frac{\Gamma(M)}{\pi^M}~\delta(\|\phi\|^2-1).
\end{equation}
We can now express the CDF of $s_{k\bar{k}}$ using the isotropic random variable ${\mbox{\boldmath $\phi$}}$. Specifically,
\begin{equation}\label{eq20}
F_{s_{k\bar{k}}}(x)=Pr\{s_{k\bar{k}}\leq x\}=Pr\{\|{\mbox{\boldmath $\phi$}}\|^2_{x{\bf \Lambda}-{{\bf \Sigma}_{k\bar{k}}}}\geq 0\}
=\int_{\|{\mbox{\boldmath $\phi$}}\|^2_{x{\bf \Lambda}-{{\bf \Sigma}_{k\bar{k}}}}\geq 0}p({\mbox{\boldmath $\phi$}})d\phi.
\end{equation}
This is an $M$-dimensional integral over the region defined by the inequality $\|{\mbox{\boldmath $\phi$}}\|^2_{x{\bf \Lambda}-{\bf
\Sigma_{k\bar{k}}}}\geq 0$ which is difficult to evaluate. To go around this, we rewrite (\ref{eq20}) as an unconstrained integral by using the unit
step function as
\begin{equation}\label{eq21}
F_{s_{k\bar{k}}}(x)=\int_{-\infty}^{\infty} p({\mbox{\boldmath $\phi$}})\tilde{u}(\|{\mbox{\boldmath $\phi$}}\|^2_{x{\bf \Lambda}
-{{\bf \Sigma}_{k\bar{k}}}})d{\mbox{\boldmath $\phi$}}.
\end{equation}
This multidimensional integral is evaluated in Appendix A, where we show that
\begin{equation}\label{cdfskkbarexp}
F_{s_{k\bar{k}}}(x)= \sum_{i=1}^M\frac{(\lambda_i x-\sigma_{i}^{k\bar{k}})^{M-1}\tilde{u}(\lambda_i x-\sigma_{i}^{k\bar{k}})}{\Pi_{j=1,j\neq i}^{M}[(\sigma_{j}^{k\bar{k}}-\sigma_{i}^{k\bar{k}})-(\lambda_j-\lambda_i)x]}.
\end{equation}
Here $\sigma_{i}^{k\bar{k}}$ is the $i^{th}$ diagonal element in matrix ${{\bf \Sigma}_{k\bar{k}}}$ and it is given by
\begin{equation}\label{sigmakkbardef}
\sigma_{i}^{k\bar{k}}= \left \{
  \begin{array}{l@{\quad \quad}l}
    \sqrt{\frac{\lambda_k}{\lambda_{\bar{k}}}}, & {\rm for~~} i = k \\%
    \sqrt{\frac{\lambda_{\bar{k}}}{\lambda_k}}, & {\rm for~~} i = \bar{k}\\%
    0, & {\rm otherwise}.%
  \end{array}
  \right.%
\end{equation}
The CDF of $s_k$, denoted by $F_{s_k}(x)$, can be derived using the same approach and is found to be the same expression as given in (\ref{cdfskkbarexp}) with $\sigma_{i}^{k\bar{k}}$ set to 1 for $i=k$ and 0 otherwise. Thus, $F_{s_k}(x)$ can be shown to be
\begin{equation}\label{cdfsk}
F_{s_{k}}(x)= \frac{\left(\lambda_k x-1\right)^{M-1}\tilde{u}\left(\lambda_k x-1\right)}{\Pi_{j=1,j\neq k}^{M}\left[-1-(\lambda_j-\lambda_k)x\right]}
+\sum_{i=1,i\neq k}^M\frac{\lambda_i^{M-1} x^{M-1}\tilde{u}(\lambda_i x)}{\left[1-(\lambda_k-\lambda_i)x\right]\Pi_{j=1,j\neq k,i}^{M}[(\lambda_i-\lambda_j)x]}.
\end{equation}
\subsection{Moments of $s_k$}\label{firstmomskkbar}
From (\ref{sk}), it is easy to see that $s_{k}$ is a positive random variable whose support is the interval $\left[0,\frac{1}{\lambda_k}\right]$. Thus, by employing integration, the first moment of $s_{k}$ can be evaluated as 
\begin{equation}\label{firstmomformula}
E[s_{k}]=\int_{0}^{\infty}\left(1-F_{s_{k}}(x)\right) dx=\int_{0}^{\frac{1}{\lambda_k}}\left(1-F_{s_{k}}(x)\right).
\end{equation}

Using (\ref{cdfskkbarexp}) and (\ref{firstmomformula}) and applying partial fraction expansion, we can show after tedious but straight forward algebraic manipulations that

\begin{equation}
E[s_k]=\frac{1}{\lambda_k}-\sum_{i=1,i\neq k}^M p_i\left(\frac{1}{\lambda_k}-\frac{\mbox{ln}\left(\lambda_{ik}\right)}{(\lambda_i-\lambda_k)}\right),\label{finalskmom1}
\end{equation}
where $\mbox{ln}(.)$ represents the natural logarithm while $\lambda_{ij}$ and $p_i$ are defined as
\begin{equation}
\lambda_{ij}\stackrel{\triangle}{=}\frac{\lambda_i}{\lambda_j},\forall~ i,j~~~\mbox{and}~~~p_i\stackrel{\triangle}{=}\frac{\lambda_i^{M-1}}{\Pi_{j=1,j\neq i}^{M}(\lambda_i-\lambda_j)}, \forall~ i.
\end{equation}
Now, observe that $\sum_{i=1}^M p_i=1$. Thus, we can use $1-\sum_{i=1,i\neq k}^M p_i=p_k$. As a result, the expression for $E[s_k]$ in (\ref{finalskmom1}) is simplified to
\begin{equation}
E[s_k]=\frac{p_k}{\lambda_k}+\sum_{i=1,i\neq k}^M p_i\frac{\mbox{ln}\left(\lambda_{ik}\right)}{(\lambda_i-\lambda_k)}.\label{finalskmom11}
\end{equation}
In a similar manner, we can express the second moment of $s_k$ as
\begin{equation}\label{skmom2}
E[s^2_{k}]=\int_{0}^{\frac{1}{\lambda_k}}2x\left(1-F_{s_k}(x)\right) dx.
\end{equation}
Thus, together with CDF expression in (\ref{cdfskkbarexp}), we get the following expression
\begin{equation}
E[s_k^2]=\frac{1}{\lambda_k^2}-\sum_{i=1,i\neq k}^Mp_i\left(\frac{1}{\lambda_k^2}-\frac{2}{\lambda_k(\lambda_i-\lambda_k)}
+\frac{2\mbox{ln}\left(\lambda_{ik}\right)}{(\lambda_i-\lambda_k)^2}\right).\label{finalskmom2}
\end{equation}
Finally, by using $1-\sum_{i=1,i\neq k}^M p_i=p_k$, we find
\begin{equation}
E[s_k^2]=\frac{p_k}{\lambda_k^2}+\sum_{i=1,i\neq k}^M\frac{2p_i}{\lambda_k(\lambda_i-\lambda_k)} -\sum_{i=1,i\neq k}^M\frac{2p_i\mbox{ln}\left(\lambda_{ik}\right)}{(\lambda_i-\lambda_k)^2}.\label{finalskmom22}
\end{equation}
\subsection{Second Moment of $s_{k\bar{k}}$}\label{secondmomskkbar}
We only need the second moment of $s_{k\bar{k}}$. We evaluate it using a similar approach that was used to evaluate the second moment of $s_k$. Specifically, we use the fact that $s_{k\bar{k}}$ has support over $\left[0,\frac{1}{\sqrt{\lambda_k\lambda_{\bar{k}}}}\right]$. Thus, after some tedious algebraic manipulations the second moment of $s_{k\bar{k}}$ can be shown to be
\begin{eqnarray}
E[s_{k\bar{k}}^2]&=&\int_{0}^{\frac{1}{\sqrt{\lambda_k\lambda_{\bar{k}}}}}2x\left(1-F_{s_{k\bar{k}}}(x)\right) dx,\nonumber\\
&=&\frac{p_k}{\lambda_k\lambda_{\bar{k}}}-\sum_{i=1,i\neq k, \bar{k}}^Mp_i\Bigg(\frac{2c_{ki}+2c_{\bar{k}i}}{\sqrt{\lambda_k\lambda_{\bar{k}}}}+\mbox{ln}\left(\lambda_{ki}\right)^{2c_{ki}\bar{\sigma}_{ki}}+\mbox{ln}
\left(\lambda_{\bar{k}i}\right)^{2c_{\bar{k}i}\bar{\sigma}_{\bar{k}i}}\Bigg),\label{finalsmom2}
\end{eqnarray}
where $\bar{\sigma}_{ji}$, $c_k$ and $c_{\bar{k}i}$ are defined as
\begin{equation}
\bar{\sigma}_{ji}=\frac{\sigma_{j}^{k\bar{k}}}{(\lambda_i-\lambda_j)},~~c_{ki}=\frac{(-\bar{\sigma}_{ki})^{M-1}}{(\bar{\sigma}_{\bar{k}i}-\bar{\sigma}_{ki})}~~\mbox{and}~~c_{\bar{k}i}=\frac{(-\bar{\sigma}_{\bar{k}i})^{M-1}}{(\bar{\sigma}_{ki}-\bar{\sigma}_{\bar{k}i})}.
\end{equation}
Thus, using the expression for $\bar{\sigma}_{ji}$, the expressions for $c_k$ and $c_{\bar{k}i}$ can be reformulated as
\begin{equation}
c_{ki}=\frac{(-1)^{M-1}\lambda_{k\bar{k}}^{(M-2)/2}}{(\lambda_i-\lambda_k)^{M-2}
\left[1-\frac{\lambda_{\bar{k}k}^2(\lambda_i-\lambda_k)}{(\lambda_i-\lambda_{\bar{k}})}\right]},~~~\mbox{and}~~~
c_{\bar{k}i}=\frac{(-1)^{M-1}\lambda_{\bar{k}k}^{(M-2)/2}}{(\lambda_i-\lambda_{\bar{k}})^{M-2}
\left[1-\frac{\lambda_{k\bar{k}}^2(\lambda_i-\lambda_{\bar{k}})}{(\lambda_i-\lambda_k)}\right]}.
\end{equation}  
Finally, by utilizing the above expressions, we can express $E[s_{k\bar{k}}^2]$ as 
\begin{eqnarray}
E[s_{k\bar{k}}^2]&=&\frac{p_k}{\lambda_k\lambda_{\bar{k}}}
-\frac{2(-1)^{M-1}\lambda_{k\bar{k}}^{(M-2)/2}}{\sqrt{\lambda_k\lambda_{\bar{k}}}}\Bigg\{\sum_{i=1,i\neq k, \bar{k}}^M \frac{p_i(1+\mbox{ln}\left(\lambda_{ki}\right))}{(\lambda_i-\lambda_k)^{M-2}
\left[1-\frac{\lambda_{\bar{k}k}^2(\lambda_i-\lambda_k)}{(\lambda_i-\lambda_{\bar{k}})}\right]}\nonumber\\
&&\hspace{4cm}+\sum_{i=1,i\neq k, \bar{k}}^M \frac{p_i(1+\mbox{ln}\left(\lambda_{\bar{k}i}\right))}{(\lambda_i-\lambda_{\bar{k}})^{M-2}
\left[1-\frac{\lambda_{k\bar{k}}^2(\lambda_i-\lambda_{\bar{k}})}{(\lambda_i-\lambda_k)}\right]}\Bigg\}.\label{finalsmom22}
\end{eqnarray}
\subsection{The pdf and second moment of $r$}\label{pdfr}
Consider the random variable $r$ defined in (\ref{zrdef}). Its CDF can
be evaluated using the approach outlined above for the CDF of $s_k$ and $s_{k\bar{k}}$ and is found to be (derivation is omitted for brevity)
\begin{equation}
F_{r}(x)=\sum_{m=1}^{M}\frac{\lambda_m^M}{|\Lam|\Pi_{i=1,i\neq m}^{M}\left(\lambda_{mi}-1\right)}e^{\frac{-1}{\lambda_m x}}~\tilde{u}(x).\label{CDFrk}
\end{equation}
Consequently, the pdf of $r$ is given by
\begin{equation}
p_r(x)=\sum_{m=1}^{M}\frac{\lambda_m^{M-1}}{x^2|\Lam|\Pi_{i=1,i\neq m}^{M}\left(\lambda_{mi}-1\right)}e^{\frac{-1}{\lambda_m x}}.\label{pdfr}
\end{equation}
We can use the above pdf to show that 
\begin{eqnarray}
E[r^2]&=&\sum_{m=1}^{M}\frac{\lambda_m^{M-1}}{|\Lam|\Pi_{i=1,i\neq m}^{M}\left(\lambda_{mi}-1\right)}\int_{0}^{\infty}\frac{e^{\frac{-1}{\lambda_m x}}}{x^2} dx=\sum_{m=1}^{M}\frac{\lambda_m^{M}}{|\Lam|\Pi_{i=1,i\neq m}^{M}\left(\lambda_{mi}-1\right)}.\label{secMomr}
\end{eqnarray}
\subsection{The pdf and second moment of $z_k$}
The second moment of $z_k$ is evaluated in this section by first deriving its pdf. However, we cannot derive the pdf of $z_k$ directly and so we derive the conditional pdf of $z_k$ conditioned on $a_k$ defined as
\begin{equation}
a_k=\sum_{m=1,m\neq k}^{M}|u(m)|^2
\end{equation}
Thus, we can rewrite $z_k$ as
\begin{equation}\label{zredf}
z_k=\frac{(1/\lambda_k)|u(k)|^2+1}{|u(k)|^2+a_k}=\frac{||{\bf u}||^2_{\bar{\bf H}_k}+1}{||{\bf u}||^2_{{\bf H}_k}+a_k},
\end{equation}
where $\bar{\bf H}_k$ and ${\bf H}_k$ are the $M\times M$ matrices with all elements set to 0 except the $k^{th}$ element on
the diagonal which is set to $\frac{1}{\lambda_k}$ and 1, respectively. By applying an approach similar to the one used with $s_k$, we can show that 
\begin{equation}\label{condpdfzk}
f_{z_k|a_k}(x)=\left[\frac{a_k}{(\lambda_kx-1)}+\frac{\lambda_k(1-a_k x)}{(\lambda_kx-1)^2}\right]e^{\frac{-(1-a_k x)}{(\lambda_kx-1)}}~  \left[\tilde{u}\left(x-\frac{1}{\lambda_k}\right)- \tilde{u}\left(x-\frac{1}{a_k}\right)\right],
\end{equation}
which allows us to evaluate the conditional moment $E[z_k^2|a_k]$ as 
\begin{equation}
E[z_k^2|a_k]=\int_{0}^{\infty}x^2f_{z_k|a_k}(x)dx=\int_{\frac{1}{\lambda_k}}^{\frac{1}{a_k}}x^2\frac{(\lambda_k-a_k)}{(\lambda_kx-1)^2}e^{\frac{-(1-a_k x)}{(\lambda_kx-1)}}dx.
\end{equation}
Here, we have utilized the fact that the support of $z_k$ is between $\mbox{min}\left(\frac{1}{\lambda_k},\frac{1}{a_k}\right)$ and $\mbox{max}\left(\frac{1}{\lambda_k},\frac{1}{a_k}\right)$. Now, by applying change of variable and using partial fraction expansion, the above integral is found to be
\begin{equation}\label{condmomzk}
E[z_k^2|a_k]=\frac{1}{\lambda_k^2}+\frac{2(\lambda_k-a_k)}{\lambda_k^3}~e^{\frac{a_k}{\lambda_k}}~\Gamma\left(0,\frac{a_k}{\lambda_k}\right)
+\frac{(\lambda_k-a_k)^2}{\lambda_k^4}~e^{\frac{a_k}{\lambda_k}}~\Gamma\left(-1,\frac{a_k}{\lambda_k}\right),
\end{equation}
where $\Gamma(\alpha,x)\stackrel{\triangle}{=}\int_x^{\infty}t^{\alpha-1}e^{-t}dt$ is the {\it Incomplete Gamma function}. The unconditional moment $E[z_k^2]$ is obtained by averaging the conditional moment
$E[z_k^2|a_k]$ over the pdf of $a_k$. The pdf of $a_k$ can be obtained using the approach of random variable $r$ and is found to be
\begin{equation}
f_{a_k}(a_k)=\frac{\lambda_k}{|\Lam|}\sum_{m=1,m\neq k}^{M}\frac{e^{-\frac{a_k}{\lambda_m}}}{\Pi_{l=1,l\neq k,m}^{M}\left(\frac{1}{\lambda_l}-\frac{1}{\lambda_m}\right)}.\label{pdfak}
\end{equation}
Thus, by using (\ref{condmomzk}) and (\ref{pdfak}), the moment $E[z_k^2]$ is found to be
\begin{align}\label{uncondmomz}
E[z_k^2]&=\sum_{m=1,m\neq k}^{M}\frac{\mbox{ln}(\lambda_m)}{|\Lam|\Pi_{l=1,l\neq k,m}^{M}\left(\frac{1}{\lambda_l}-\frac{1}{\lambda_m}\right)}\nonumber\\
&+\sum_{m=1,m\neq k}^{M}\frac{\lambda_m\Big(1+_2F_1\left(1,1;2;\eta_{mk}\right)-~_2F_1\left(1,1;3;\eta_{mk}\right)\Big)}{\lambda_k|\Lam|\Pi_{l=1,l\neq k,m}^{M}\left(\frac{1}{\lambda_l}-\frac{1}{\lambda_m}\right)}\nonumber\\
&+\sum_{m=1,m\neq k}^{M}\frac{\lambda_m^2}{\lambda_k^2|\Lam|\Pi_{l=1,l\neq k,m}^{M}\left(\frac{1}{\lambda_l}-\frac{1}{\lambda_m}\right)}\left(\frac{1}{3}~_2F_1\left(1,2;4;\eta_{mk}\right)-~_2F_1\left(1,2;3;\eta_{mk}\right)\right),
\end{align}
where $\eta_{mk}=1-\lambda_{mk}$ and $~_2F_1\left(\alpha,\beta;\gamma;x\right)$ is the {\it Hypergeometric Function} defined as
\begin{equation}\label{hyperdef}
~_2F_1\left(\alpha,\beta;\gamma;x\right)\stackrel{\triangle}{=}\frac{1}{B(\beta,\gamma-\beta)}\int_0^1x^{\beta-1}(1-x)^{\gamma-\beta-1}(1-xz)^{-\alpha}dx,
~~\mbox{Re}(\gamma)>\mbox{Re}(\beta)>0,
\end{equation}
and $B(x,y)$ is the {\it Beta Function} defined as
\begin{equation}\label{Betadef}
B(x,y)\stackrel{\triangle}{=}\int_0^1t^{x-1}(1-t)^{y-1}dt,~~~~~~~~\mbox{Re}(x)>0,~~\mbox{Re}(y)>0.
\end{equation}

\section{Simulation Results}\label{sim}
\begin{figure}[H]
\centering
\includegraphics[width=0.45\textwidth]{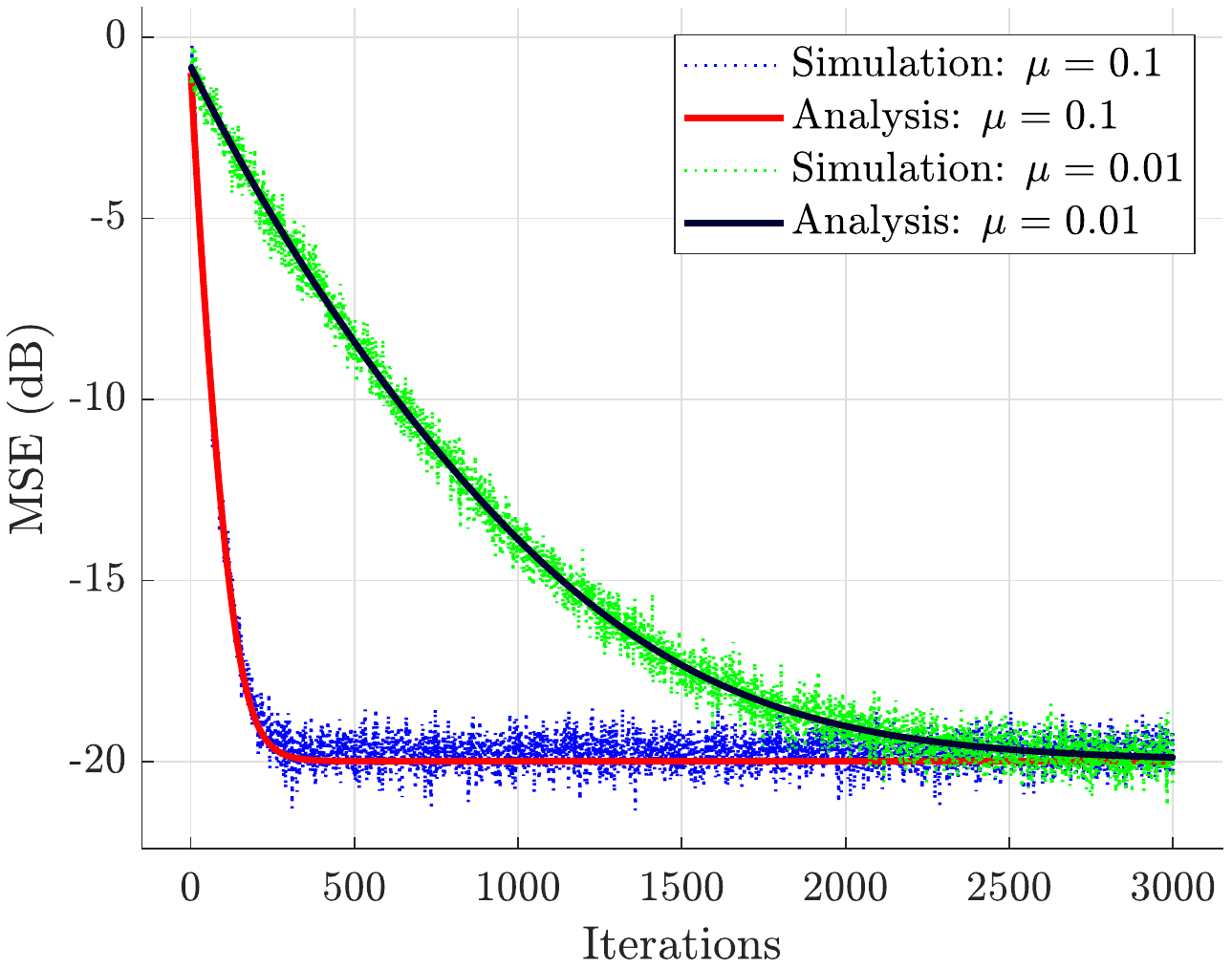}
\caption{Analytical and simulation results for the MSE learning curves of the NLMS algorithm for $\alpha_c=0.5$.}
\label{fig:MSE1}
\end{figure}

In this section, the performance analysis of the NLMS algorithm is investigated for an unknown system identification problem. The system noise is a zero mean i.i.d. sequence with variance 0.01 which sets the SNR to 20 dB. Throughout the simulation, the adaptive filter used has the same length as that of the unknown system. The optimal weight vector is ${\bf{w}^o}=[0.227,~0.460,~0.688,~0.460,~0.227]^T$ as we assume $\bq_i=\bzero,~\forall~i$. The input to the adaptive filter and unknown system is correlated complex Gaussian input which is generated with correlation matrix with entries ${\bf R}(i,j)=\alpha_c^{|i-j|}$ with correlation factor\footnote{The case $\alpha_c=0$ corresponds to the white case while $\alpha_c=1$ corresponds to the fully correlated case} $\alpha_c$ ($0<\alpha_c<1$). The numerical results are averaged over $100$ independent runs. The objective of our simulations is to validate the derived analytical results for both steady-state and transient analysis. In order to validate the derived analytical result for the transient behavior, the MSE of the NLMS algorithm is analyzed in Fig.~\ref{fig:MSE1} for two different values of step-size which are 0.1 and 0.01. The value of correlation factor $\alpha_c$ is set to 0.5. It can be easily seen that the results validate our theoretical findings for both step-sizes.

\section{Conclusion}\label{con}
We analyzed the NLMS adaptive filter driven by correlated circularly symmetric Gaussian input.
The paper demonstrates that the MSE performance can be articulated in terms of the moments of ratios of quadratic forms in isotropic
random variables. The CDFs of these ratios then obtained using novel complex integration techniques, which are then used to evaluate
the corresponding moments in closed form expressions. Our MSE results apply to
the transient, stead-state, and tracking results. Our proposed approach for the analysis of the NLMS could be extended other normalized adaptive filters (e.g. Recursive Least Square (RLS) and affine projection algorithms).
\section*{Appendix A: Evaluating the CDF of $s_{k\bar{k}}$}
In order to evaluate the integral in (\ref{eq21}), we utilize the following integral representations of the step and the delta functions \cite{isotropic}
\begin{equation}\label{stepdelta}
\tilde{u}(x)=\frac{1}{2\pi}\int_{-\infty}^{\infty} \frac{e^{x(j\omega+\beta)}}{(j\omega+\beta)} d\omega~~\mbox{and}~~
\delta(x)=\frac{1}{2\pi}\int_{-\infty}^{\infty} e^{x(\alpha+j\omega)} d\omega,
\end{equation}
These are valid for any $\alpha,~\beta >0$ which are free parameters that we can conveniently choose \cite{Ryzhik}. After replacing the delta and the step functions with their equivalent integral representations, the pdf $p({\mbox{\boldmath $\phi$}})$
and $\tilde{u}(\|{\mbox{\boldmath $\phi$}}\|^2_{x{\bf \Lambda}-{{\bf \Sigma}_{k\bar{k}}}})$ in (\ref{eq21}) can be rewritten as
\begin{align}
p({\mbox{\boldmath $\phi$}})&=\frac{\Gamma(M)}{2\pi^{M+1}}\int_{-\infty}^{\infty} e^{(\|{\mbox{\boldmath $\phi$}}\|^2-1)(\alpha+j\omega)} d\omega,
\end{align}
and
\begin{align}
\tilde{u}(\|{\mbox{\boldmath $\phi$}}\|^2_{x{\bf \Lambda}-{{\bf \Sigma}_{k\bar{k}}}})&=\frac{1}{2\pi}\int_{-\infty}^{\infty} \frac{e^{(\|{\mbox{\boldmath $\phi$}}\|^2_{x{\bf \Lambda}-{{\bf \Sigma}_{k\bar{k}}}})(j\omega+\beta)}}{(j\omega+\beta)} d\omega,
\end{align}
which allows us to set up the CDF of $s_{k\bar{k}}$ in (\ref{eq21}) as
\begin{equation}\label{CDFintegrals}
F_{s_{k\bar{k}}}(x)= \frac{\Gamma(M)}{4\pi^{M+2}}~e^{-\alpha}\int_{-\infty}^{\infty}
\int_{-\infty}^{\infty}\int_{-\infty}^{\infty} e^{-{\mbox{\boldmath $\phi$}}^{*}(-\alpha {\bf I}+({{\bf \Sigma}_{k\bar{k}}}-x{\bf
\Lambda})(j\omega_1+\beta)-j\omega_2{\bf I}){\mbox{\boldmath $\phi$}}} d{\mbox{\boldmath $\phi$}}~ e^{-j\omega_2}d\omega_2 \frac{1}{j\omega_1+\beta}d\omega_1.
\end{equation}
Next, we first focus on the inner integral
\begin{equation}\label{multidimI}
\int_{-\infty}^{\infty} e^{-{\mbox{\boldmath $\phi$}}^{*}(-\alpha {\bf I}+({{\bf \Sigma}_{k\bar{k}}}-x{\bf
\Lambda})(j\omega_1+\beta)-j\omega_2{\bf I}){\mbox{\boldmath $\phi$}}} d{\mbox{\boldmath $\phi$}}.
\end{equation}
By inspecting the above, we note that it is similar to the Gaussian density integral with a covariance $\Sigma_c$ given by
\begin{equation}
\Sigma_{c}=(-\alpha {\bf I}+({{\bf \Sigma}_{k\bar{k}}}-x{\bf \Lambda})(j\omega_1+\beta)-j\omega_2{\bf I})^{-1}.
\end{equation}
Note that $\Sigma_{c}$ is complex. Hence, $\alpha$ should be chosen to make the real part of $\Sigma_{c}$ positive definite. Thankfully, the final result will not depend on the value of $\alpha$ we choose. Here, we are just making an analogy to the Gaussian integral (see \cite{ISIT} for a formal proof) which gives
\begin{equation}
\frac{1}{\pi^M}\int_{-\infty}^{\infty} e^{-{\mbox{\boldmath $\phi$}}^{*}(-\alpha {\bf I}+({{\bf \Sigma}_{k\bar{k}}}-x{\bf
\Lambda})(j\omega_1+\beta)-j\omega_2{\bf
I}){\mbox{\boldmath $\phi$}}} d{\mbox{\boldmath $\phi$}}=\frac{1}{\Big|-\alpha {\bf I}+({{\bf \Sigma}_{k\bar{k}}}-x{\bf
\Lambda})(j\omega_1+\beta)-j\omega_2{\bf I}\Big|}.
\end{equation}
Consequently, the CDF will take the following form
\begin{equation}
F_{s_{k\bar{k}}}(x)= \frac{\Gamma(M)}{4\pi^{2}}~e^{-\alpha}\int_{-\infty}^{\infty}
\int_{-\infty}^{\infty}\frac{e^{-j\omega_2}}{\Big|-\alpha {\bf I}+({{\bf \Sigma}_{k\bar{k}}}-x{\bf
\Lambda})(j\omega_1+\beta)-j\omega_2{\bf I}\Big|}d\omega_2~\frac{1}{(j\omega_1+\beta)}d\omega_1.
\end{equation}
At this stage, we focus on the second inner integral with respect to $\omega_2$. To evaluate this integral, we use partial fraction
expansion to represent the determinant as
\begin{align}
&\frac{1}{\Pi_{i=1}^M(-\alpha+(\sigma_{i}^{k\bar{k}}-\lambda_{i}x)(j\omega_1+\beta)-j\omega_2)}\nonumber\\
&~~~~~~~~~~~~~~~~~~~~~~~~=\frac{1}{(j\omega_1+\beta)^{M-1}}\sum_{i=1}^M\frac{\eta_i}{(-\alpha+(\sigma_{i}^{k\bar{k}}-\lambda_{i}x)(j\omega_1+\beta)-j\omega_2)},
\end{align}
where constant $\eta_i$ is given by
\begin{equation}\label{etadef}
\eta_i=\frac{1}{\Pi_{j=1,j\neq i}^{M}[(\sigma_{j}^{k\bar{k}}-\sigma_{i}^{k\bar{k}})-(\lambda_j-\lambda_i)x]},
\end{equation}
and $\sigma_{i}^{k\bar{k}}$ is defined in (\ref{sigmakkbardef}). We can now use residue theory to evaluate the integral with respect to $\omega_2$. Specifically, we use the following integral solutions \cite{Ryzhik}
\begin{align}
\int_{-\infty}^{\infty}(\beta+ix)^{-\nu}e^{-ipx}dx=\frac{2\pi (-p)^{\nu-1}e^{\beta p}}{\Gamma(\nu)}~\tilde{u}(-p),\label{firstsol}
\end{align}
and
\begin{align}
\int_{-\infty}^{\infty}(\beta-ix)^{-\nu}e^{-ipx}dx=\frac{2\pi p^{\nu-1}e^{-\beta p}}{\Gamma(\nu)}~\tilde{u}(p),\label{secondsol}%
\end{align}
which is valid for $\mbox{Re}(\nu)>0$ and $\mbox{Re}(\beta)>0$. Thus, by employing the above relations, solution of the integral with respect to $\omega_2$ is found to be
\begin{equation}
\frac{1}{2\pi}\int_{-\infty}^{\infty}\frac{e^{-j\omega_2}}{\Big|-\alpha {\bf I}+({{\bf \Sigma}_{k\bar{k}}}-x{\bf
\Lambda})(j\omega_1+\beta)-j\omega_2{\bf
I}\Big|}d\omega_2=\frac{1}{(j\omega_1+\beta)^{M-1}}\sum_{i=1}^M\eta_ie^{\alpha-(\sigma_{i}^{k\bar{k}}-\lambda_i x)(j\omega_1+\beta)},
\end{equation}
Eventually, the CDF of $s_{k\bar{k}}$ can be set up as follows
\begin{eqnarray}
F_{s_{k\bar{k}}}(x)
&=&\frac{\Gamma(M)}{2\pi}\sum_{i=1}^M\eta_i\int_{-\infty}^{\infty}\frac{1}{(j\omega_1+\beta)^M}e^{-(\sigma_{i}^{k\bar{k}}-\lambda_i x)(j\omega_1+\beta)}d\omega_1
\end{eqnarray}
Finally, after again using the residue theory and (\ref{etadef}), the CDF of $s_{k\bar{k}}$ given in (\ref{cdfskkbarexp}) is
found.

%
%
%
%
%


\begin{thebibliography}{17}
\bibitem{sayed}
A. H. Sayed, {\em Fundamentals of Adaptive Filtering}, New York: Wiley-Interscience, 2003.

\bibitem{NLMS}
{J. I., Nagumo and A. Noda},
\newblock ``A learning method for system identification,"
\newblock {\em {IEEE Transactions on Automatic Control}}, vol 12, pp.~282-287, 1967.

\bibitem{TranDataNon}
{T. Y. Al-Naffouri and A. H. Sayed},
\newblock ``Transient analysis of data normalized adaptive filters," \newblock {\em {IEEE Transactions on
Signal Processing}}, vol 51, No. 3, pp.~639-652, March, 2003.

\bibitem{Douglas}
{S. C. Douglas and T. H. Y. Meng},
\newblock ``Normalized Data Nonlinearities for LMS Adaptation," \newblock {\em {IEEE Transactions on
Signal Processing}}, vol 42, No. 6, pp.~1352-1365, June, 1994.

\bibitem{Barrault}
{G. Barrault, M. H. Costa, J. C. M. Bermudez and A. Lenzi},
\newblock ``A new analytical model for the NLMS algorithm," \newblock {\em {Proc. IEEE ICASSP}}, vol. 4, pp.~41-44, March, 2005.

\bibitem{Lobato}
{E. M. Lobato, O. J. Tobias, and R. Seara},
\newblock ``Stochastic model for the NLMS algorithm with correlated Gaussian data," \newblock {\em {Proc. IEEE ICASSP}}, vol. 3, pp.~760-763. 2006.

\bibitem{CostaBermudaz}
{M. H. Costa and J. C. M. Bermudaz},
\newblock ``An improved model for the normalized LMS algorithm with gaussian inputs and large number of coefficients," \newblock {\em{Proc. IEEE ICASSP}}, vol. 2, pp. 1385-1388, 2002.

\bibitem{Slock}
{D. T. M. Slock},
\newblock ``On the convergence behavior of the LMS and the normalized LMS algorithms," \newblock {\em {IEEE Transactions on
Signal Processing}}, vol 41, No. 9, pp.~2811-2825, Sept., 1993.

\bibitem{Ali}
{A. Ali, M. Moinuddin, and T. Y. Al-Naffouri},
\newblock ``The NLMS Is Steady-State Schur-Convex," \newblock {\em {IEEE Signal Processing Letters}}, vol. 28, pp. 389-393, 2021.


\bibitem{Tarrab}
{M. Tarrab, and A. Feuer},
\newblock ``Convergence and Performance Analysis of the Normalized LMS algorithm with Uncorrelated Gaussian Data," \newblock {\em {IEEE Transactions on Information Theory}}, vol 34, No. 4, pp.~680-691, July, 1988.



\bibitem{NLMSmean}
{T. Y. Al-Naffouri, M. Moinuddin, and M. S. Sohail},
\newblock ``Mean Behavior of the NLMS Algorithm for Correlated Gaussian Inputs," \newblock {\em {IEEE Signal Processing Letters}}, vol. 18, Issue 1, pp 7-10, Jan. 2011.

\bibitem{Mathai}
{A. M. Mathai and S. B. Provost},
\newblock ``Quadratic Forms in Random Variables,"
\newblock {\em {New York: Marcel Dekker}}, 1992.

\bibitem{Johnson}
{N. L. Johnson and S. Kotz},
\newblock ``Continuous Univariate Distributions,"
\newblock {\em {New York: Houghton Mifflin}}, vol 2, 1970.

\bibitem{isotropic}
{B. Hassibi and T. L. Marzetta},
\newblock ``Multiple-antennas and isotropically random unitary inputs: The received signal density in closed-form," \newblock {\em {IEEE Transactions on Information Theory}}, vol. 48, No. 6, pp. 1473-1484, June 2002.

%
%



\bibitem{Ryzhik}
{I. S. Gradshteyn, I. M. Ryzhik},
\newblock ``Table of integrals, series, and products, corrected and enlarged edition", \newblock {\em {Academic Press, INC, New York}}, 1980.

%
%

\bibitem{ISIT}
{T.~Y. Al-Naffouri and B.~Hassibi},
\newblock ``{On the distribution of indefenite quadratic forms in Gaussian
  random variables},''
\newblock {\em {International Symposium on Information Theory (ISIT), South
  Korea}}, 2009.
 

%
%
%
%
%
%

\end{thebibliography}
\end{document}